\documentclass[a4paper,12pt]{article}
\usepackage{latexsym}
\usepackage{amssymb}
\usepackage{epsfig}
\date{}

\title{Effect of detrending on multifractal characteristics}

\author{P.~O\'swi\c ecimka$^{1,*}$, S.~Dro\.zd\.z$^{1,2}$, J.~Kwapie\'n$^1$ and A.Z. G\'orski$^{1}$}

\begin{document}
\maketitle
\noindent
$^1$\textit{Institute of Nuclear Physics, Polish Academy of Sciences, Krak\'ow, Poland}
$^2$\textit{Faculty of Physics, Mathematics and Computer Science, Cracow University of Technology, Krak\'ow, Poland}

\begin{abstract}

Different variants of MFDFA technique are applied in order to investigate various (artificial and real-world) time series. Our analysis shows that the calculated singularity spectra are very sensitive to the order of the detrending polynomial used within the MFDFA method. The relation between the width of the multifractal spectrum (as well as the Hurst exponent) and the order of  the polynomial used in calculation is evident. Furthermore, type of this relation itself depends on the kind of analyzed signal. Therefore, such an analysis can give us some extra information about the correlative structure of the time series being studied.
\end{abstract}

\vskip0.5cm

{\it * corresponding author; e-mail: Pawel.Oswiecimka@ifj.edu.pl}

\section{Introduction}

In recent years investigation of complex systems with regard to their fractal properties has become one of the
elementary methods of such systems analysis~\cite{kwapien2012}. Multifractal structures were identified in systems from
various areas such as physics~\cite{muzy08,oswiecimka06,oswiecimka05,subramaniam08},
biology~\cite{ivanov99,makowiec09,rosas02}, chemistry~\cite{stanley88,udovichenko02},
economics~\cite{drozdz10,oswiecimka08,zhou09} and even music~\cite{voss75,hsu90,jafari07,su06,oswiecimka11}. A simple
consequence of this fact is both popularization of the existing methods of multifractal analysis and proposing new, even
more advanced methods. Among the problems the multifractal analysis faces is coping with a trend present in analysed
data~\cite{wu07}. This is because typically only the signal fluctuations have to be considered. In such a case, a trend
should be eliminated by using one of the detrending methods, which is best suited for a particular kind of the trend.
The simplest and most popular way of detrending is to remove the trend with a definite functional form. However,
identifying precise functional form of the trend becomes very difficult or even impossible if the data is nonstationary.
Moreover, the trend can itself depend on the considered time span. Thus, in practice it often happens that detrending is
nothing else but taking the functional form ad hoc and subtracting it from the analysed signal. Unfortunately, this can
lead to spurious findings and their erroneous interpretation. 

One of the most popular methods of the multifractal analysis is Multifractal Detrended Fluctuation Analysis
(MFDFA)~\cite{kantelhart2002,drozdz09,makowiec2010}. Within its framework, the supposed trend is removed from time
series before multifractal spectra are calculated. The trend is represented by a polynomial with chosen degree. However,
as it was mentioned above, a choice of the detrending polynomial order is crucial. Using a polynomial of too high order
can result in suppressing part of the low-frequency fluctuations which are erroneously identified as a part of the
trend. On the other hand, polynomial of too low order does not eliminate non-stationarity sufficiently. This problem was
discussed in Ref.\cite{kantelhart2001} where several types of trends and their influence on fractal characteristics of
data were considered. 

In this paper, we systematically study the multifractal characteristics of different signals by means of the MFDFA method with different orders of the detrending polynomials. We consider both the mathematical multifractals and the multifractal time series coming from real-world observables or generated by a computer. The diversity of the considered signals ensures that our study is comprehensive and not restricted to one type of correlations only. The primary multifractal characteristics we apply here is the singularity spectrum and its derivatives.

\section{Methodology}

The MFDFA procedure was proposed by Kantelhardt {\it et al.} in ~\cite{kantelhart2002} and is one of the most frequently applied algorithms of calculating the multifractal spectra. Its popularity owes to, among other, the simplicity of its implementation and the reliability of obtained results in the case of non-stationary time series. In this method, one calculates the signal profile first:
\begin{equation}
Y\left(j\right) =\sum_{i=1}^j[x_{i}-<x>]\ \  j=1...N ,
\end{equation}
where $< >$ denotes averaging over time series. Then this profile is divided into $2M_s$ disjoint segments $\nu$ of
length $s$ starting both from the beginning and the end of the series. For each box, the trend is estimated by fitting a
polynomial $P^{(m)}_{\nu}()$ of order $m$. Next, this trend is subtracted and variance of the detrended data is
calculated inside each segment:
\begin{equation}
F^{2}(\nu,s)=\frac{1}{s}\Sigma_{k=1}^{s}\lbrace Y((\nu-1)s+k)-P^{(m)}_{\nu}(k)\rbrace
\end{equation}
Finally, the $q$th-order fluctuation function is derived according to the equation:
\begin{equation}
F_q(s)=\lbrace\frac{1}{2M_{s}}\Sigma_{\nu=1}^{2M_{s}}[F^{2}(\nu,s)]^{q/2}\rbrace^{1/q},
\end{equation}
where \textit{q} can take any real value except zero. In this paper, we restrict $q$ within the range $-4\leq q \leq 4$. This procedure is repeated for different segments' lengths $s$. If the analysed signal is fractal, then $F_q$ scales within some range of $s$ according to a power law:
\begin{equation}
F_q\sim s^{h(q)},
\label{wzor1}
\end{equation}
where $h(q)$ denotes the generalized Hurst exponent. For a monofractal signal, $h(q)$ is independent of $q$ and equals the Hurst exponent $h(q)=H$. On the other hand, for a multifractal time series, $h(q)$ is decreasing function of $q$ and the ordinary Hurst exponent is obtained for $q=2$. The multifractal spectrum can be calculated by means of the following relation:
\begin{equation}
\alpha = h(q)+qh^{'}(q) \quad \hbox{and} \quad f(\alpha)=q[\alpha-h(q)]+1 ,
\end{equation}
where $\alpha$ denotes the strength of a singularity and $f(\alpha)$ is the fractal dimension of a points set with 
particular $\alpha$. For a multifractal time series, the shape of the singularity spectrum is similar to a wide inverted
parabola. The left and right wing of the parabola refers to the positive and negative values of $q$, respectively. For pure
multifractals the lower is the value of $\alpha$ the higher moment is considered. The maximum of the spectrum is located
at $\alpha(q=0)$. For a monofractal signal, $f(\alpha)$  shrinks to one point. The richness of the multifractal is
evaluated by the width of its spectrum:
\begin{equation}
\Delta \alpha= \alpha_{max} - \alpha_{min},
\end{equation}
where $\alpha _{min}$ and $\alpha _{max}$ stand for the extreme values of $\alpha$. The larger is $\Delta \alpha$, the more rich is dynamics and the more developed is the multifractal.

It is worth mentioning that MFDFA method can be easily connected with classical multifractal formalism in which 
multiscaling properties are characterized by a partition function~\cite{kantelhart2002,muzy1994}:
\begin{equation}
Z(q,s) = \Sigma_{\nu=1}^{N(s)}\mu^{q}_{\nu}(s)
\end{equation}
where $\mu_{\nu}(s)$ stands for measure (or the box probability) covering the segment $\nu$ of size $s$ and $q$ is the
same parameter as in case of MFDFA method. The fluctuation function can be related to the partition function according
to the formula:
\begin{equation}
Z(q,s) = \Sigma_{\nu=1}^{2M_{s}}[F^{2}(\nu,s)]^{q/2}
\end{equation}
In case of fractal time series we expect scaling relation similar to Eq. \ref{wzor1}:
\begin{equation}
Z(q,s) \sim s^{\tau(q)}
\end{equation}
where $\tau(q)=qh(q)-1$ is called scaling exponent. For a monofractal time series $\tau(q)$ is linear function of $q$
whereas for
multifractal one $\tau(q)$ is nonlinear (concave) function. The singularity spectrum is given by the Legendre
transformation of
scaling exponent and expressed by the formulas:
\begin{equation}
\alpha=\tau'(q)\quad \textrm{and} \quad f(\alpha)=q\alpha-\tau(q),
\end{equation}

\section{Analysis of Fractional Brownian Motion}

\begin{figure}
\includegraphics[width=0.5\textwidth,height=0.4\textwidth]{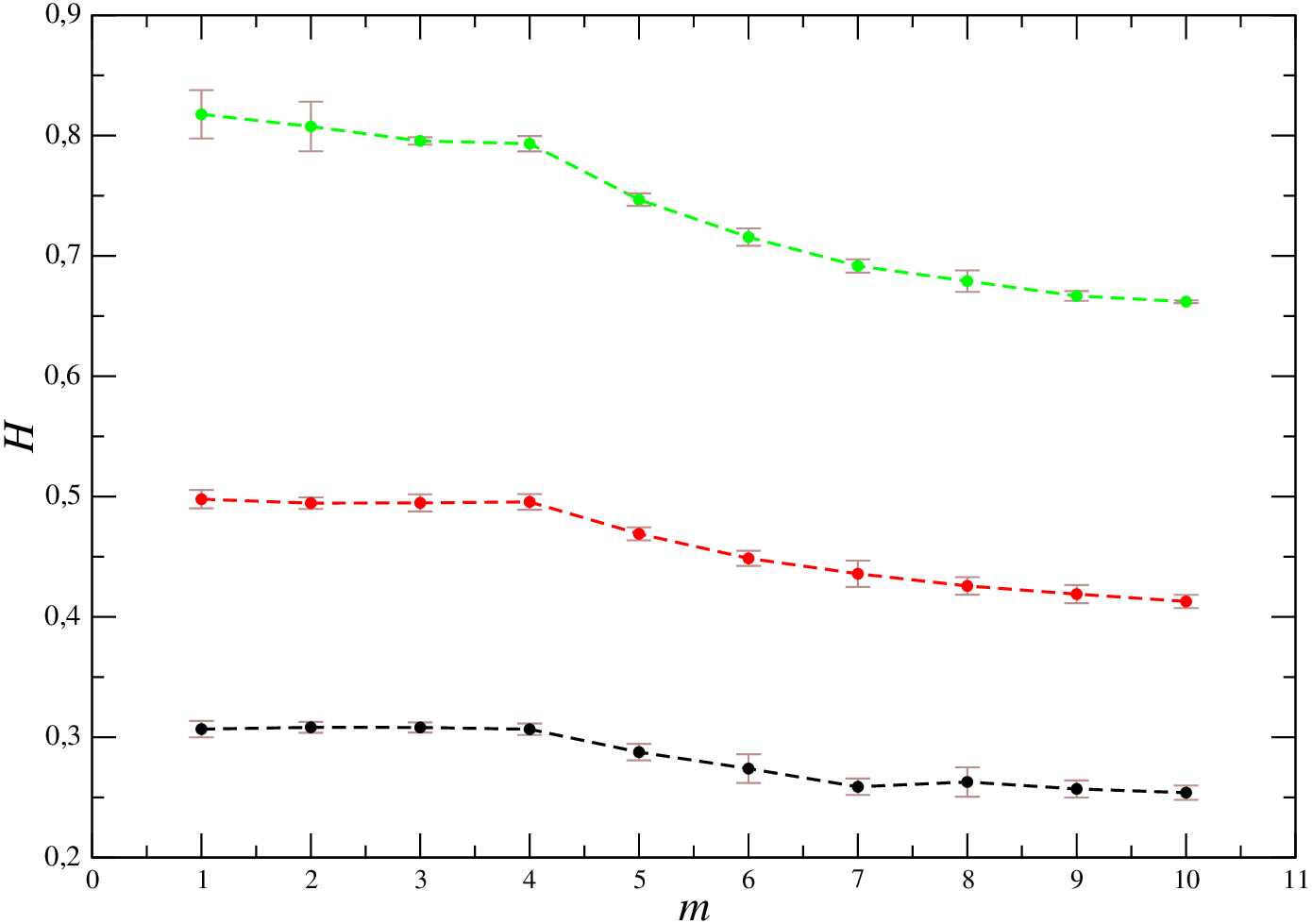}
\hspace{0.3cm}
\includegraphics[width=0.5\textwidth,height=0.4\textwidth]{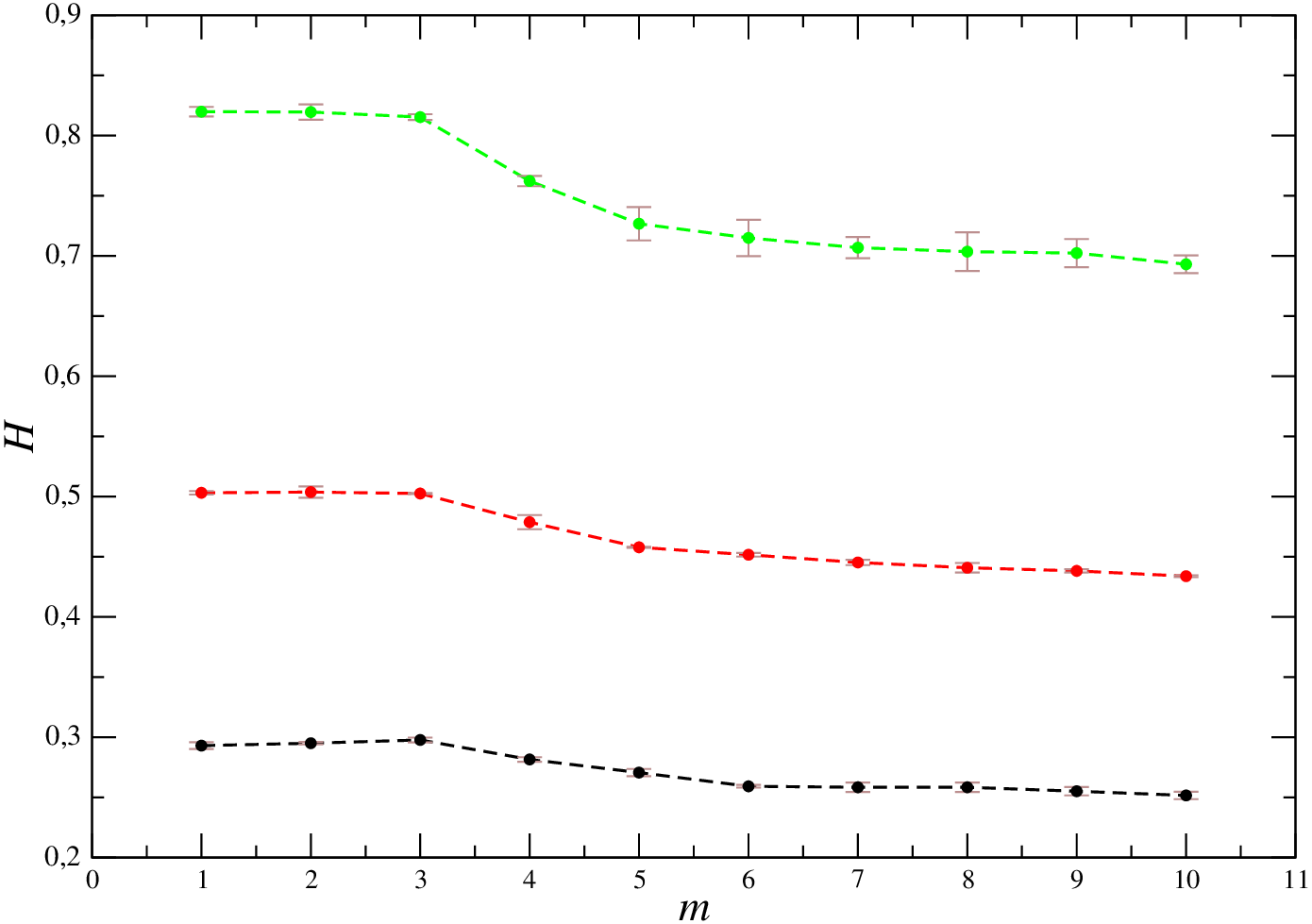}
\caption{The average Hurst exponent $H$ as a function of detrending polynomial order $m$ calculated for three artificial fractional Brownian motions: $H=0.8$ (top), $H=0.5$ (middle), and $H=0.3$ (bottom). Left panel refers to time series of 100,000 points, while right panel to time series of 1 million points. Error bars indicate standard deviation calculated from 10 independent realizations of the corresponding process.}
\label{figg1}
\end{figure}

We start our study with an analysis of a monofractal time series represented by the fractional Brownian motion 
(${\textrm B_H}$). The long-term correlations of this Gaussian processes are entirely characterized by the Hurst
exponent $H$. For $0.5<H<1$, the data is positively correlated (persistent), which means that the signal is likely to
follow the trend. ${\textrm B_H}$ with $0<H<0.5$ is negatively correlated (antipersistent) and it indicates that the
signal has a tendency to frequently change the trend direction. $H=0.5$ indicates a linearly uncorrelated time series.
We consider data of length of 100,000 and 1 million points with $H$=0.3, 0.5 and 0.8. The results for each process are
averaged over its ten independent realizations in order to be statistically significant. In each case, the fractal
analysis was performed by using MFDFA with the detrending polynomial of order $m$ in the range $(1,10)$. The results are
shown in Figure \ref{figg1}.
It can be easily observed that the calculated Hurst index depends on the polynomial order $m$ for all the considered 
time series. For small values of $m$, the estimated $H$ is the largest and, at the same time, the closest to its
theoretical value. Curiously enough, for shorter time series, $H(m)$ function is approximately constant for $m$ from 
the range $[1,4]$ and it declines for higher values of $m$. However, for longer series $H(m)$ decreases slightly for
$m=4$. This effect  is independent of kind of correlations of time series and  is result of accuracy of Hurst exponent
estimating. For longer time series, the Hurst exponent is calculated for greater number of scales then for shorter one
thus, in this case $H$ is estimated more precisely.  For $m>5$ the Hurst index rather saturates with only small
fluctuations. The difference between the extreme values of $H$ ($\Delta H=\max(H(m))-\min(H(m))$) is the largest for the
persistent signal ($\Delta H=0.125$). For the uncorrelated and antipersistent time series, $\Delta
H=0.08$. These results suggest that detrending of ${\textrm B_H}$ with a polynomial of order larger than 3 does not
retrieve the theoretical value of $H$. In this case, the detrending procedure is too effective and its results not only
remove the trend but also a part of the fluctuations. Moreover, the effectiveness of detrending depends significantly on
correlations in time series and is much better for persistent time series.

\begin{figure}
\includegraphics[width=0.5\textwidth,height=0.4\textwidth]{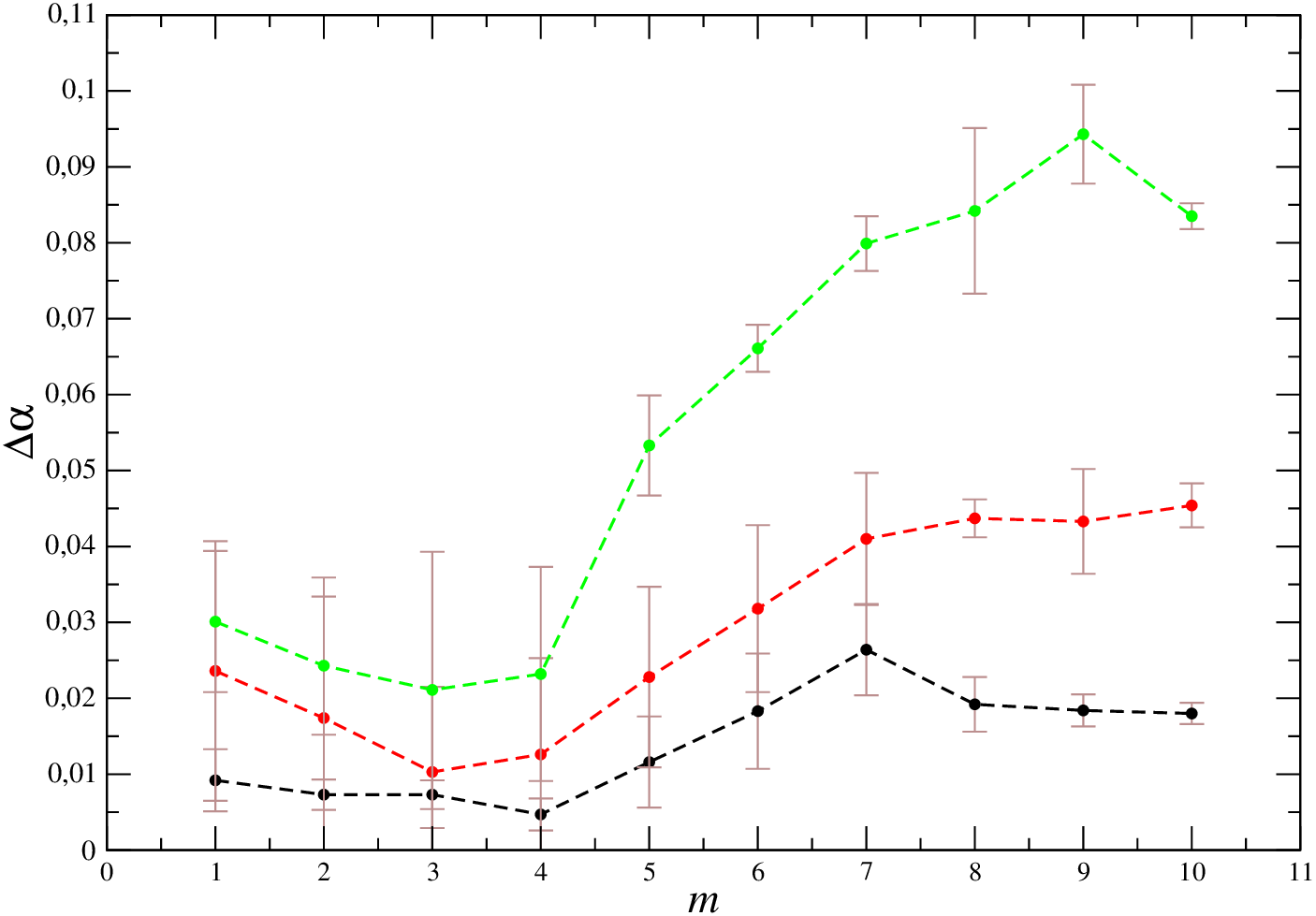}
\hspace{0.3cm}
\includegraphics[width=0.5\textwidth,height=0.4\textwidth]{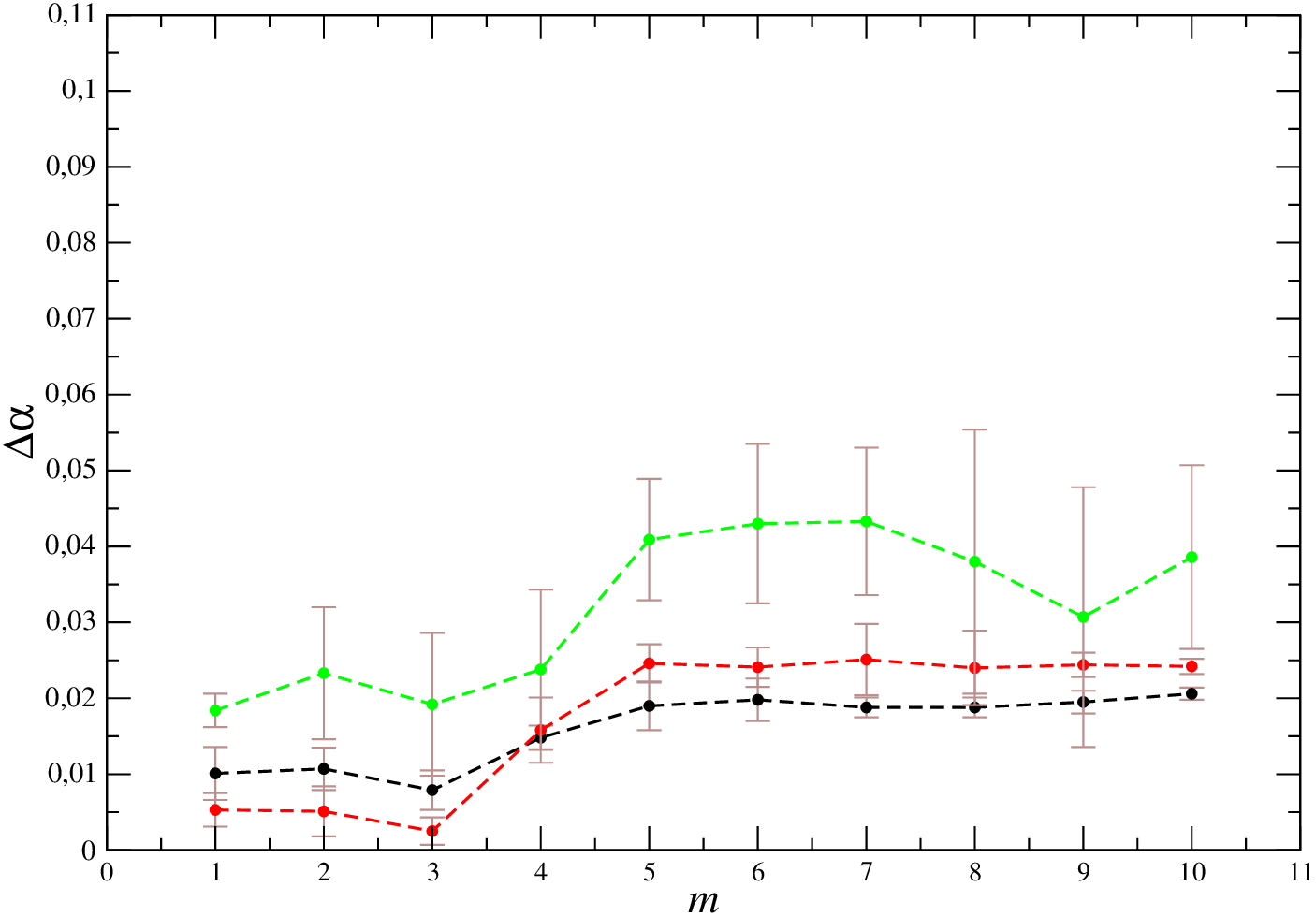}
\caption{The average width of singularity spectrum $\Delta \alpha$ as a function of detrending polynomial order $m$ 
for three artificial fractional Brownian motion signals: $H=0.8$ (top), $H=0.5$ (middle), and $H=0.3$ (bottom). Left
panel refers to time series of 100,000 points, while right panel to time series of 1 million points. Error bars indicate
standard deviation calculated from 10 independent realizations of the corresponding process.}
\label{figg2}
\end{figure}

Because the analysed time series of ${\textrm B_H}$ are monofractals, their theoretical singularity spectra are single
points with $\alpha=H$. But in practice, because of finite accuracy of the calculations, instead of a single point we
obtain a narrow spectrum~\cite{grech2012}. Keeping this property in mind, we calculate $\Delta \alpha$ for all the
investigated time series. In Figure \ref{figg2} we present how the width of the multifractal spectra changes with the
degree $m$ of the detrending polynomial. But, in contrast to the results showed above, the character of the function
$\Delta \alpha(m)$ depends on time series length. For time series with 100,000 points, $\Delta \alpha$ decreases for $1
< m < 4$. For larger $m$ and both persistent and uncorrelated time series, $\Delta \alpha$ is a monotonically increasing
function of $m$, but the rate of this increase is larger for the correlated time series. The width of the spectrum
calculated for the antipersistent data rises only for $5<m<8$ and saturates for $m > 8$. As might be expected, $\Delta
\alpha$ obtained for longer time series takes smaller values than its counterpart for shorter signals. The width of the
multifractal spectrum is approximately constant for $1<m<3$ and $5<m<10$ with the larger average values appearing for
the latter range of $m$. The widest spectra are again observed for strongly positively correlated signals what suggests
that the use of the high-order polynomials in detrending procedure affects the multifractal spectra more significant in
the case of time series with long trends. However, it should be noted that in all the cases the width of the singularity
spectra is narrow enough for the considered signals to be regarded as monofractals.

\begin{figure}
\begin{center}
\includegraphics[width=0.65\textwidth,height=0.5\textwidth]{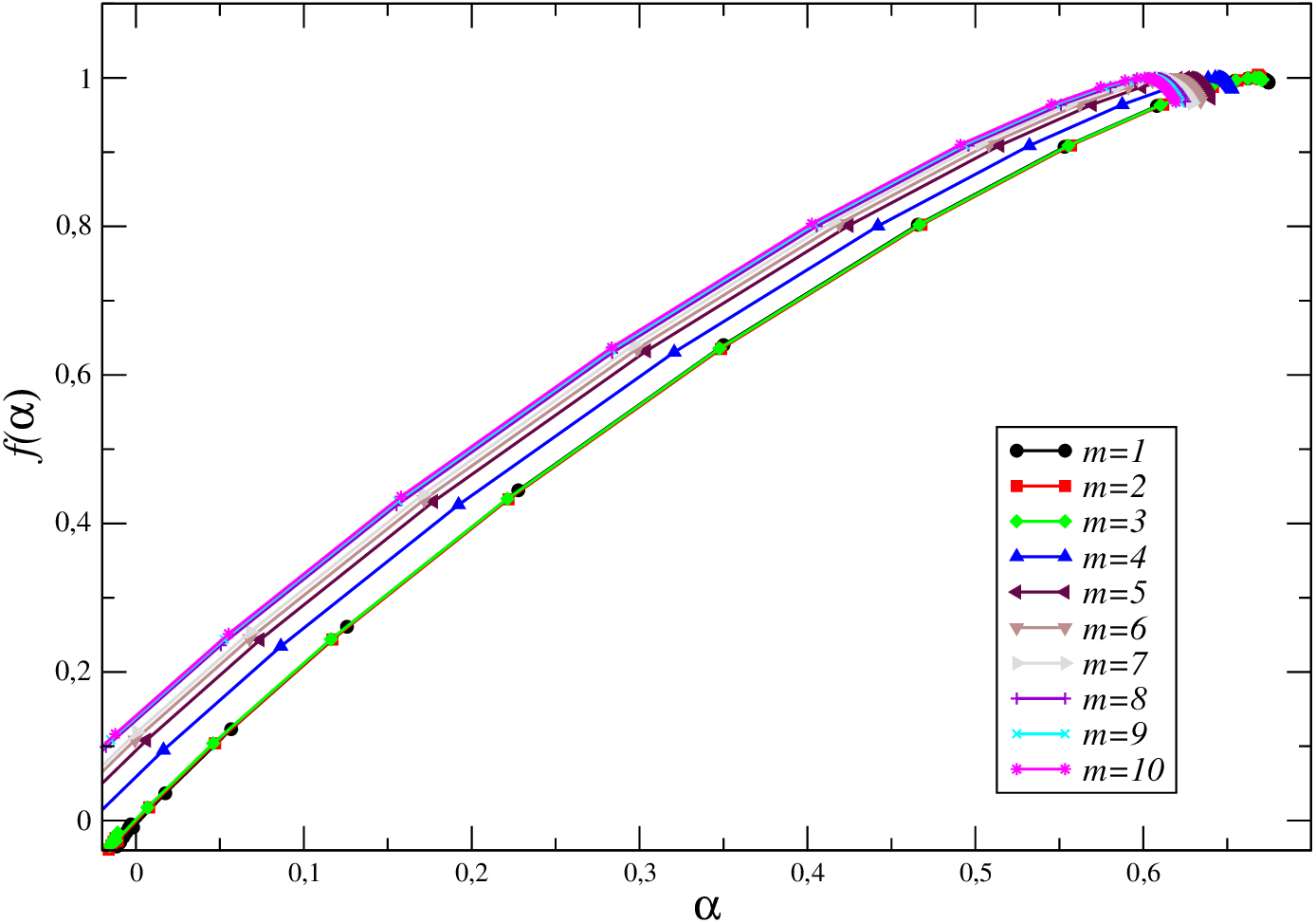}
\caption{Singularity spectra $f(\alpha)$ for the L\'evy process obtained for different polynomial orders in MFDFA.}
\label{figg3} 
\end{center}
\end{figure}

\section{Analysis of bifractal time series}

\begin{figure}
\includegraphics[width=0.5\textwidth,height=0.4\textwidth]{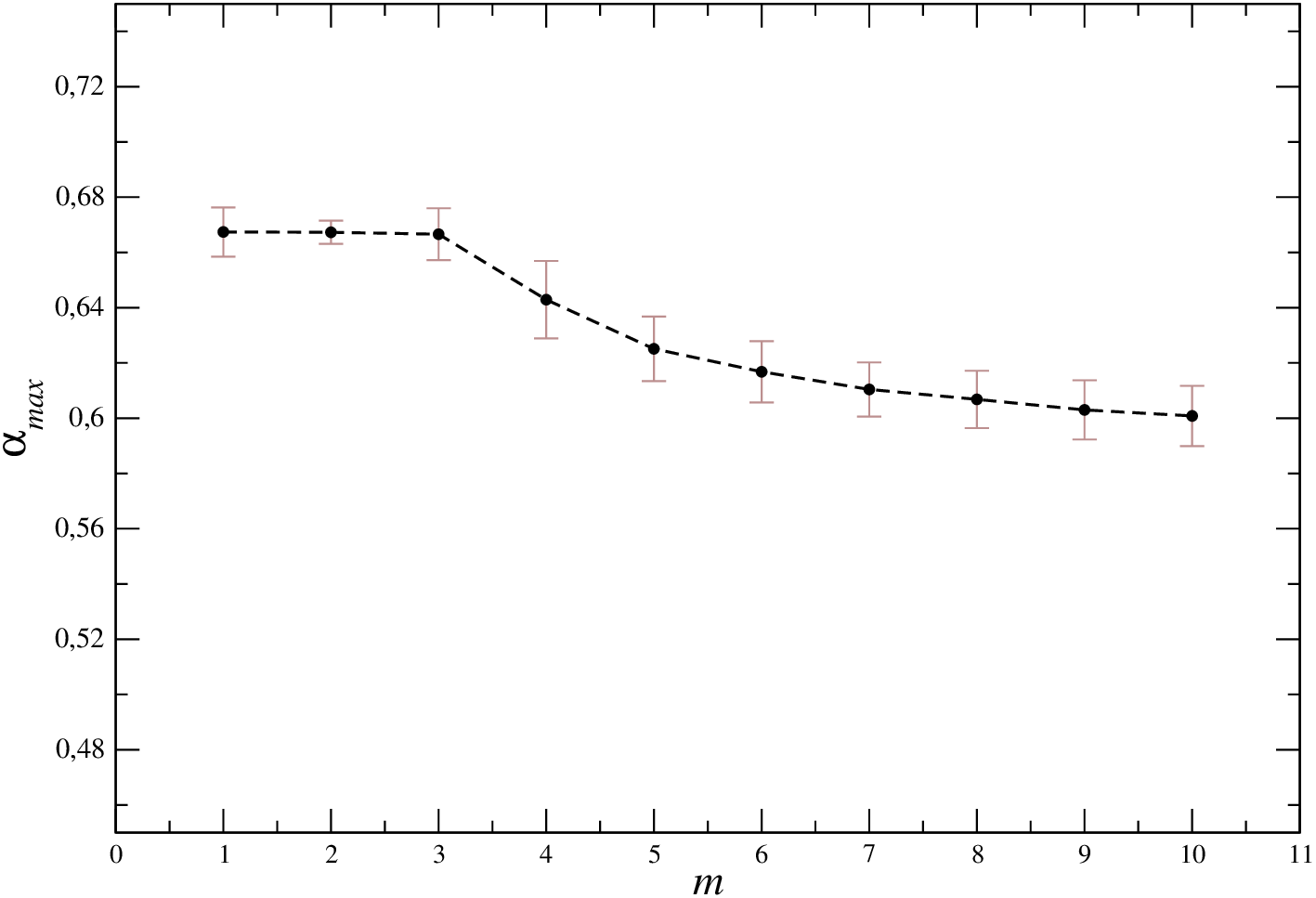}
\hspace{0.3cm}
\includegraphics[width=0.5\textwidth,height=0.4\textwidth]{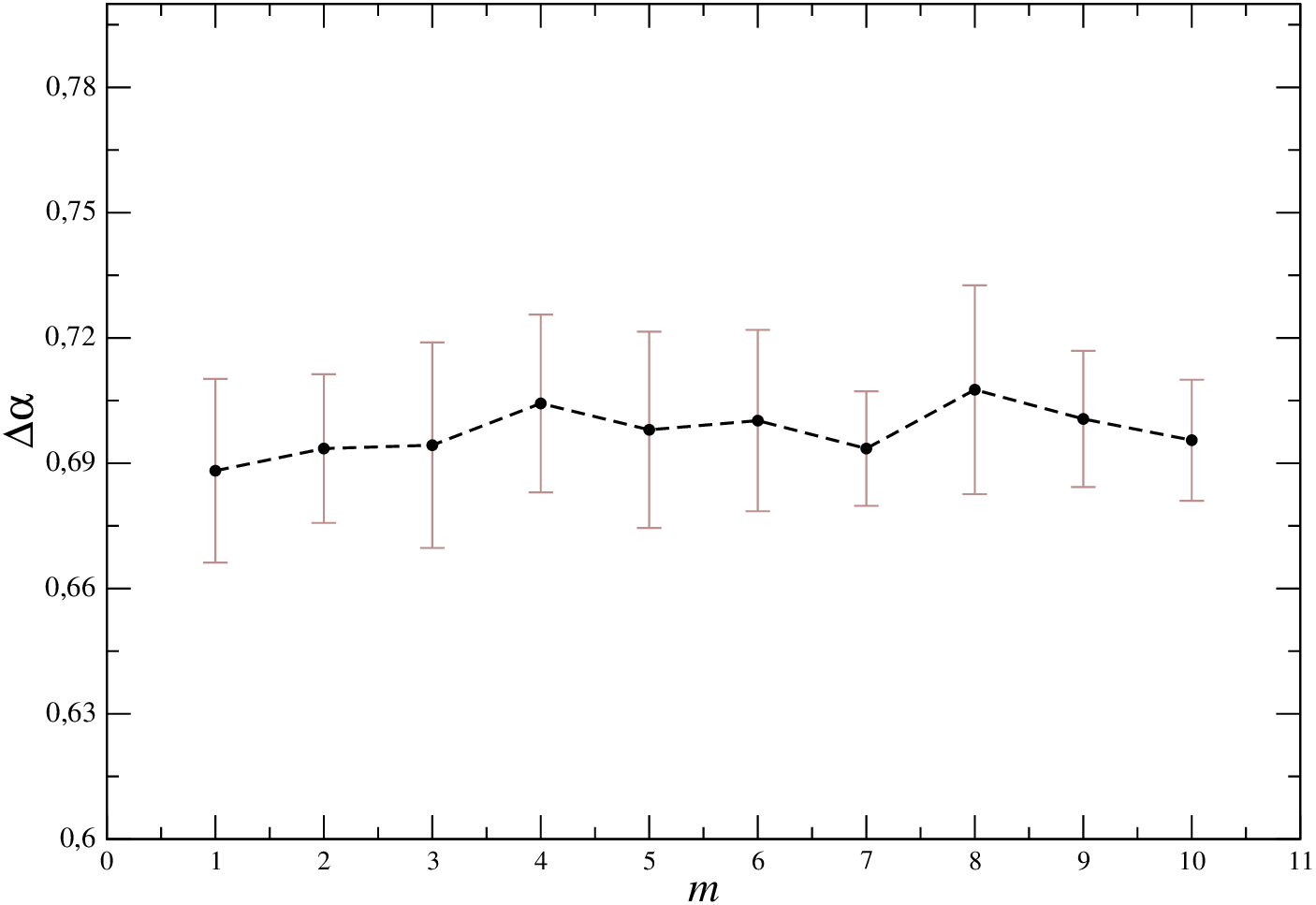}
\caption{Fractal characteristics of the L\'evy process. Left: position of the $f(\alpha)$ maximum as a function of detrending polynomial order $m$. Right: width of the singularity spectrum $\Delta \alpha$ as a function of $m$. Error bars indicate standard deviation calculated from 10 independent realizations of the corresponding process.}
\label{figg4}
\end{figure}

Another class of the processes we analyze in this paper are the L\'evy processes~\cite{nakao00}. It was shown that
the multifractal spectrum for the uncorrelated fluctuations obeying the power law distribution $P(x)\sim
x^{-(\alpha_L+1)}$ (where $\alpha_L$ is called the L\'evy index) consists of only two points and can be expressed by:

\begin{equation}
\alpha = \left\{ \begin{array}{cc} 1/\alpha_L & (q \le \alpha_L) \\
0 & (q > \alpha_L) \end{array} \right. \ \ 
f(\alpha) = \left\{ \begin{array}{cc} 1 & (q \le \alpha_L) \\
0 & (q > \alpha_L) \end{array} \right.
\end{equation}
As we can see, this processes can in some sense be considered as examples 
of a mixture of two kind of fractals. Similar to the previous case, we performed
 the analysis for time series of length of 100,000 and 1 million points but, due to similarity of the results, we
discuss only the case of 1 million points. We take the L\'evy index $\alpha_L=1.5$. In Figure \ref{figg3}, we show the
spectra $f(\alpha)$ calculated for different values of $m$. It can be easily noticed that the estimated spectra are not
single points but $f(\alpha)$ takes also the intermediate values between the points of theoretical spectrum. This effect
is known as an artifact of the MFDFA method that appears in every case of bifractal data. The presented spectra
systematically shift towards the smaller $\alpha$'s with increasing values of $m$. It is better visible in Figure
\ref{figg4} (left panel) where the average over the extrema position ($\alpha_{max}$) as a function of $m$ are depicted.
We can see that, for $1<m<3$, the $\alpha_{max}$  is equal to the theoretical value of 0.66. For a larger polynomial
degree, the position of the spectrum maximum decreases and, for $m=10$, $\alpha_{max}$ reaches the value of 0.6.
Interestingly enough, the width of the spectra is practically stable as a function of $m$ (Figure \ref{figg4}, right
panel). These results mean that, for the L\'evy processes, the fractal characteristics calculated for time series
detrending with polynomial of degree from the range $[1,3]$, reconstruct almost perfectly the theoretical spectrum.
However, for $4<m<10$ the calculated spectra suggest more a volatile signal than it is in reality, although the
dynamics of the underlying processes is not impoverished.

\section{Binomial Multifractal Cascade}

As an example of the multifractal time series we consider a binomial cascade ~\cite{oswiecimka06}. This mathematical 
model of a deterministic multifractal, by reference to binary numbers, can be defined by the following formula:
\begin{equation}
x_k=a^{n(k-1)}(1-a)^{n_{max}-n(k-1)},
\end{equation}
where $x_k$ is a time series of $2^{n_{max}}$ points $(k=1...2^{n_{max}})$, the parameter $a$, which is responsible for
the fractal properties, takes values within the range (0.5,1), and $n(k)$ denotes the number of 1's in the binary
representation of the index $k$. The fractal properties of  model are well known and quantified by the equations of the
mutifractal spectrum:
\begin{equation}
\alpha = - {1 \over \ln (2)} {{a^q\ln(a)+(1-a)^q\ln(1-a)} \over {a^q + (1-a)^q}}
\end{equation}
\begin{eqnarray}
f(\alpha) = - {q \over \ln(2)} {{a^q\ln(a) + (1-a)^q\ln(1-a)} \over {a^q + 
(1-a)^q}}-{{-\ln[a^q+(1-a)^q]} \over {\ln(2)}}.
\end{eqnarray}

\begin{figure}
\begin{center}
\includegraphics[width=0.65\textwidth,height=0.5\textwidth]{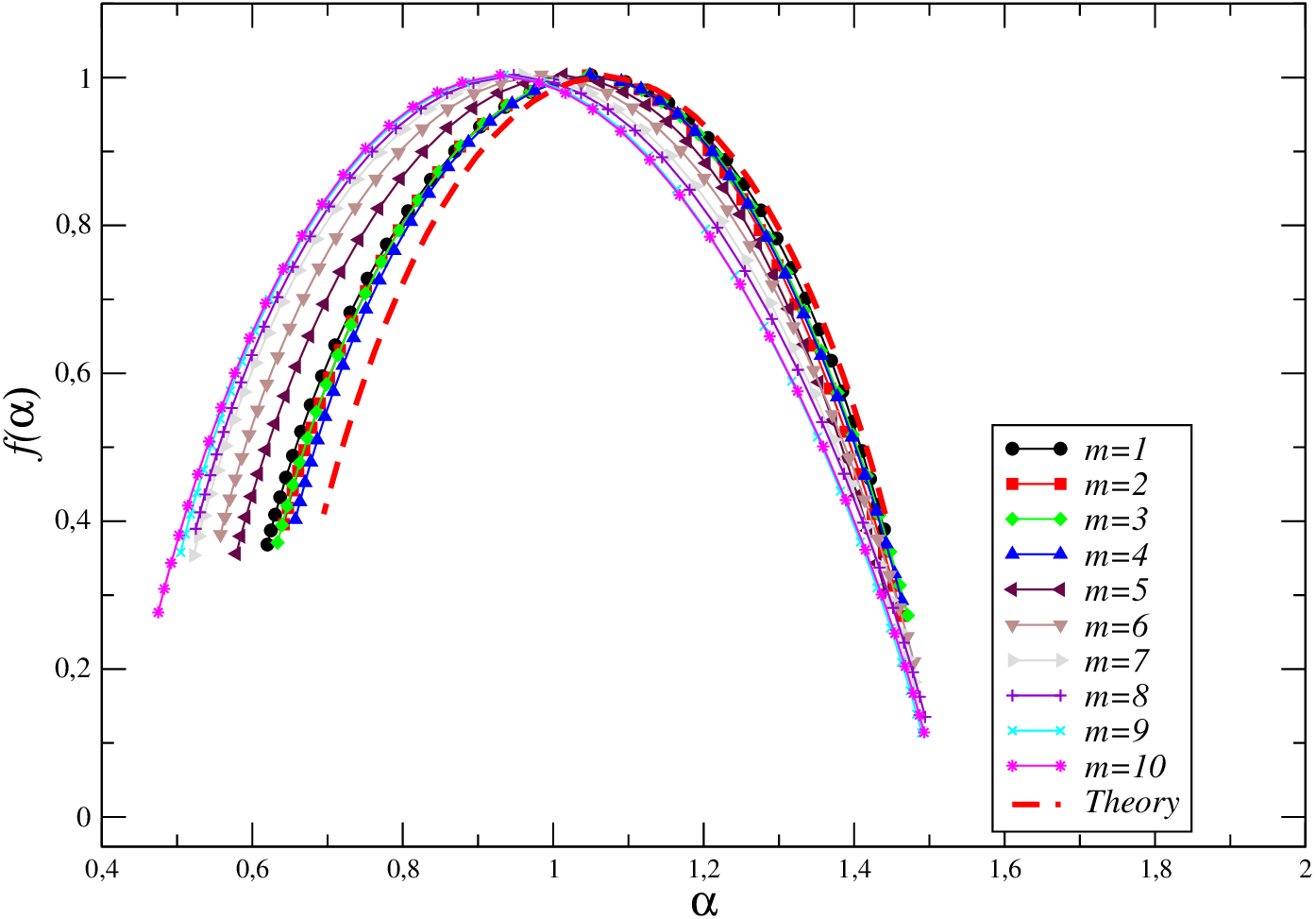}
\caption{Singularity spectra $f(\alpha)$ for binomial cascade obtained for different polynomial orders in MFDFA.}
\label{figg5}
\end{center}
\end{figure}

In our analysis, we take $a=0.65$ and $n_{max}=17$, so our time series is 131072 points long. The calculated spectra of singularities for different values of the polynomial degree used in MFDFA detrending precedure are depicted in Figure \ref{figg5}.

\begin{figure}
\includegraphics[width=0.5\textwidth,height=0.4\textwidth]{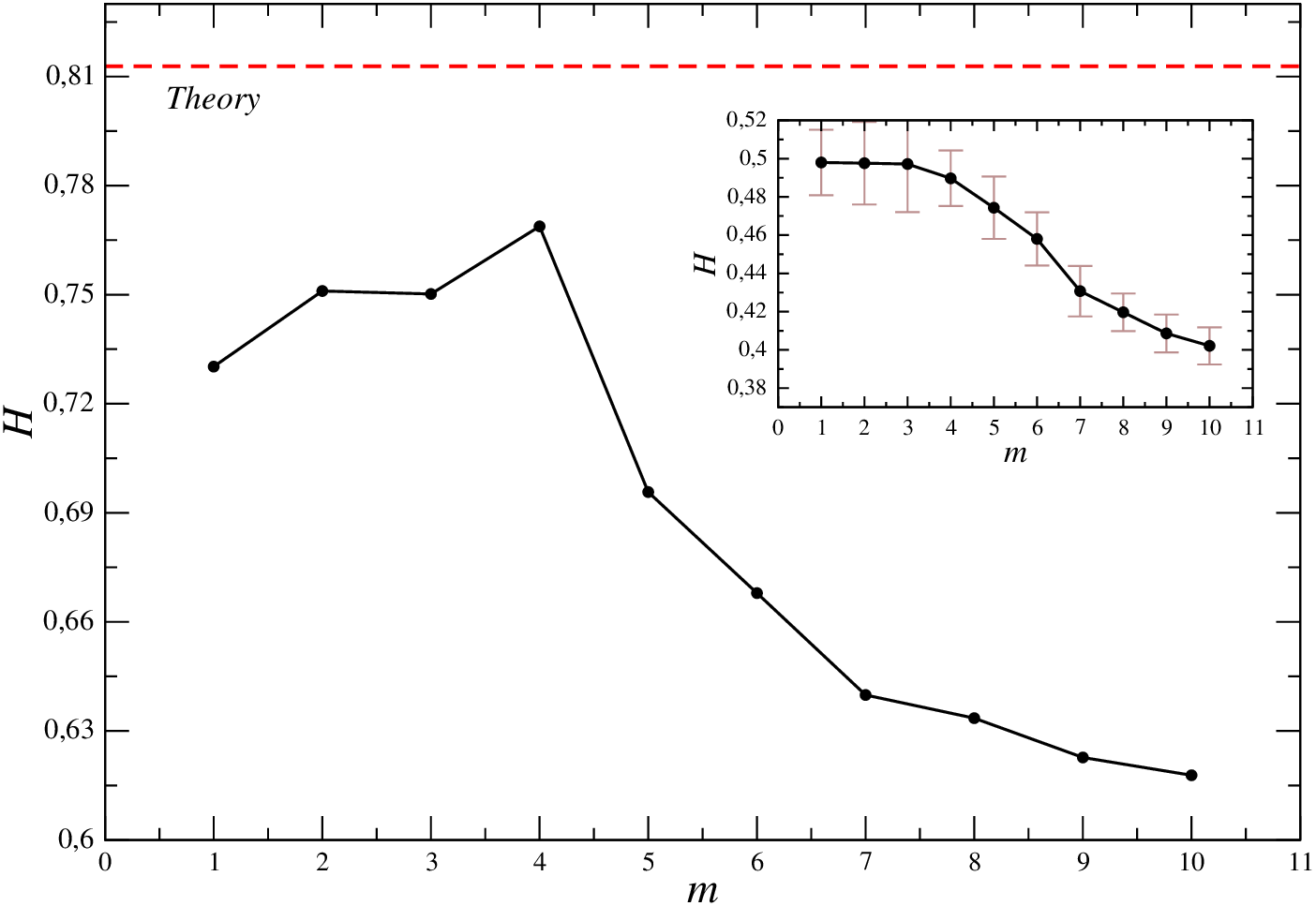}
\hspace{0.3cm}
\includegraphics[width=0.5\textwidth,height=0.4\textwidth]{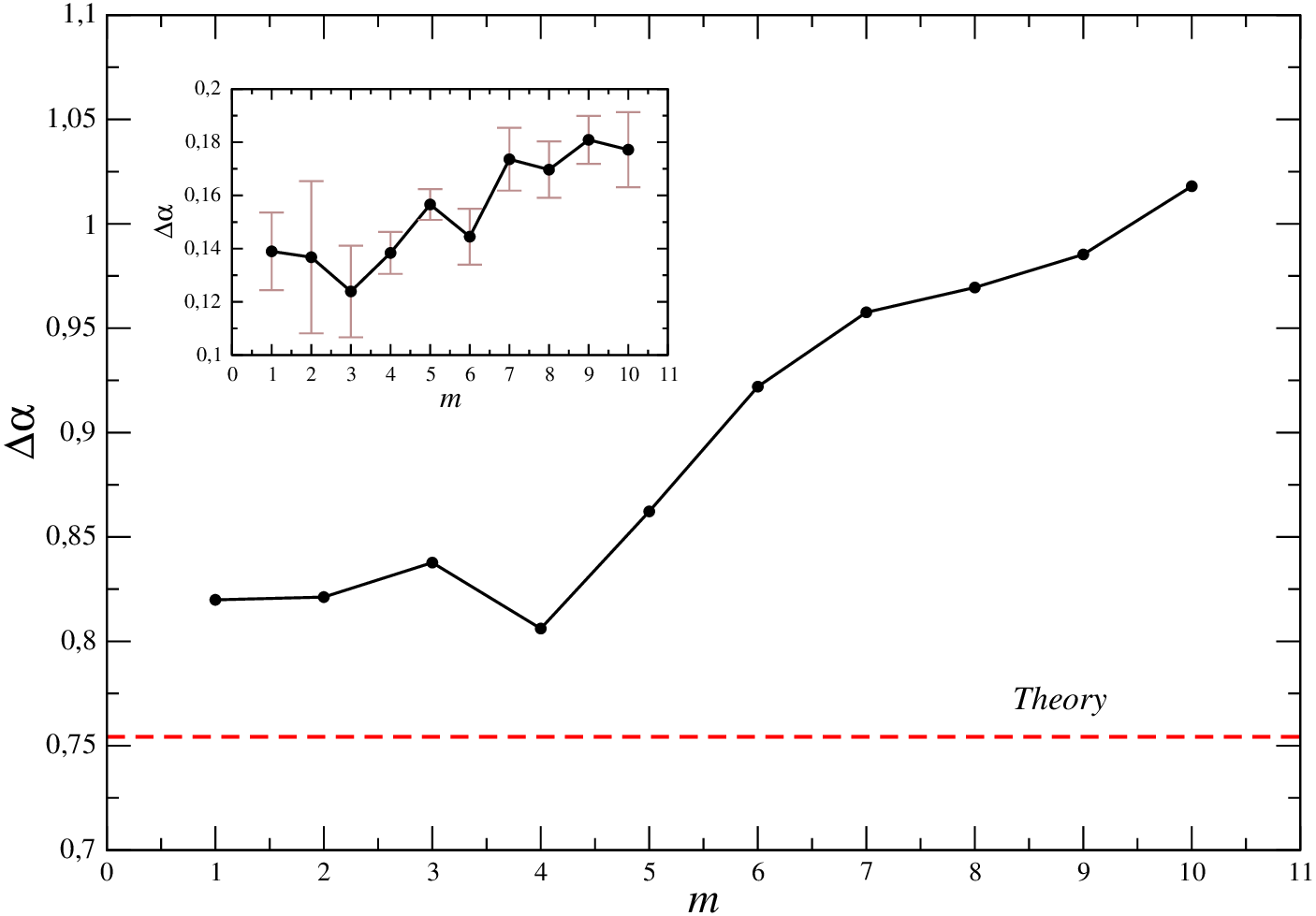}
\caption{Hurst exponent $H$ (left) and  width of the singularity spectrum $\Delta \alpha$ (right) as functions of the detrending polynomial order $m$ for binomial cascade. The inset presents the average results for the reshuffled data. Error bars indicate standard deviation calculated from 10 independent randomly shuffled time series.}
\label{figg6}
\end{figure}

In all the cases, we obtained broad spectra $f(\alpha)$ which confirm that the analyzed time series is multifractal. 
Moreover, the presented results show that, as in analysed above cases, for larger values of $m$, the spectra
systematically shift towards antipersitency. This fact is better noticeable on chart \ref{figg6} (left panel) where the
Hurst exponent as a function of $m$ is depicted. Interestingly, for small values of $m$, the $H$ index increases with
$m$, but for $4<m<10$, $H(m)$ is decreasing function of $m$. It should be noted that for all values of $m$, the
estimated Hurst exponents are smaller than their theoretical counterparts. To see the significance of the obtained
results, we performed the fractal analysis also for randomly shuffled time series. The average over ten independent
reshufflings is shown in the inset of Figure \ref{figg6} (left panel). The expected value of the Hurst exponent equal to
0.5, denoting a lack of correlations in the underlying signal, is retrieved for $1<m<3$. For larger values of $m$,
similar to the original time series, the Hurst index decreases monotonically, reaching $H=0.4$ for $m=10$.

In Figure \ref{figg6} (right panel), we present the widths of the calculated multifractal spectra. Only for small 
values of $m$, $\Delta \alpha$ is approximately constant. However, for higher polynomial degrees, $\Delta \alpha$
increases with $m$. This suggests that MFDFA variants with large $m$ show richer dynamics of the process than it is in
reality. As the inset in Figure \ref{figg6} (right panel), we show the results obtained for the reshuffled data. Also in
this case, $\Delta \alpha$ rises with $m>4$, but the rate of this increase is much slower than for the original data.

Summarizing the results depicted on Fig. \ref{figg5} and \ref{figg6}, we conclude that the multifractal spectrum
for $m=4$ is the closest to the theoretical $f(\alpha)$. This fact indicates that in the analysed
variant of binomial cascade, a polynomial of order 4 is the best approximation of the trend.

\section{Analysis of Forex market data}

\begin{figure}
\includegraphics[width=0.5\textwidth,height=0.4\textwidth]{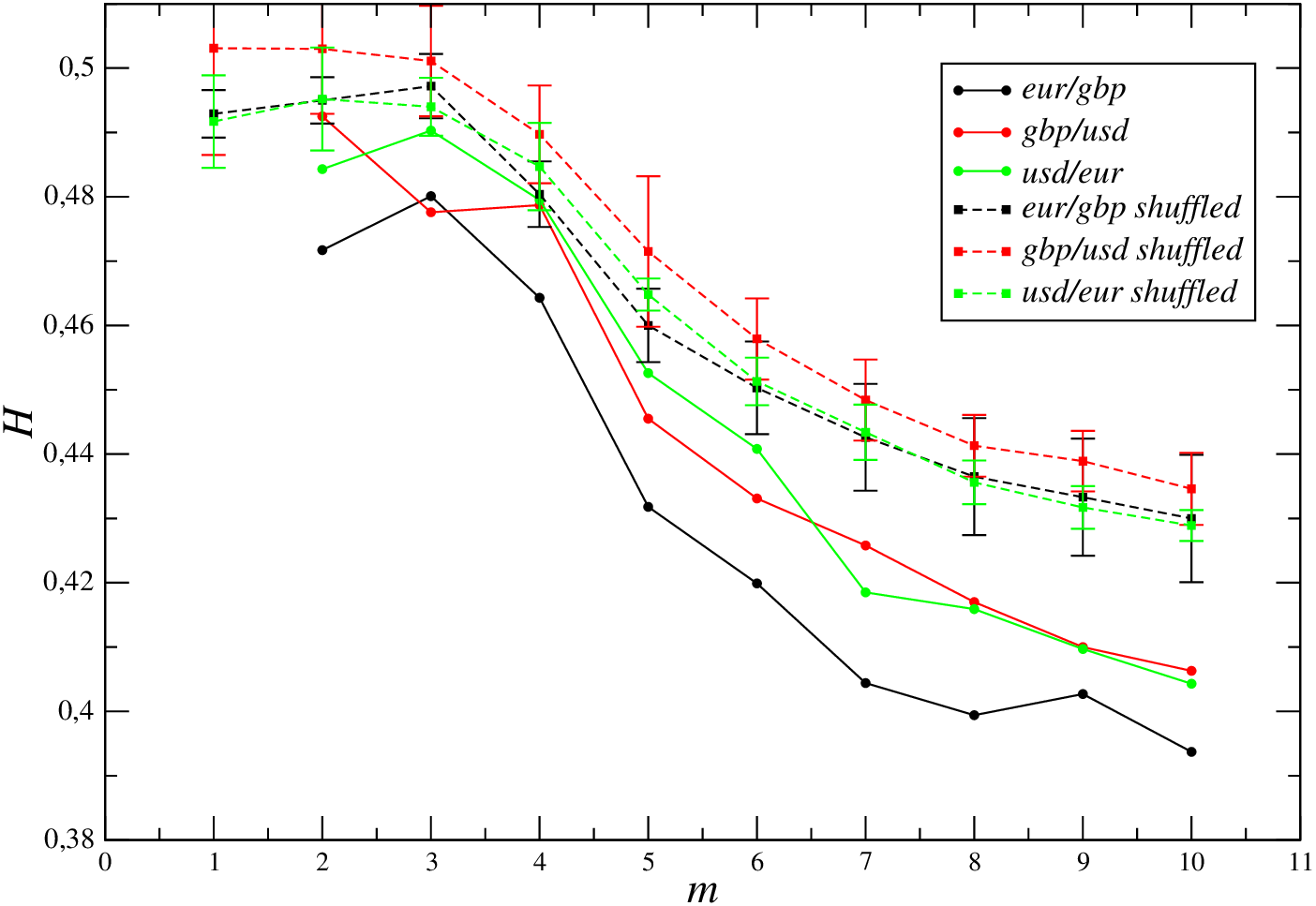}
\hspace{0.3cm}
\includegraphics[width=0.5\textwidth,height=0.4\textwidth]{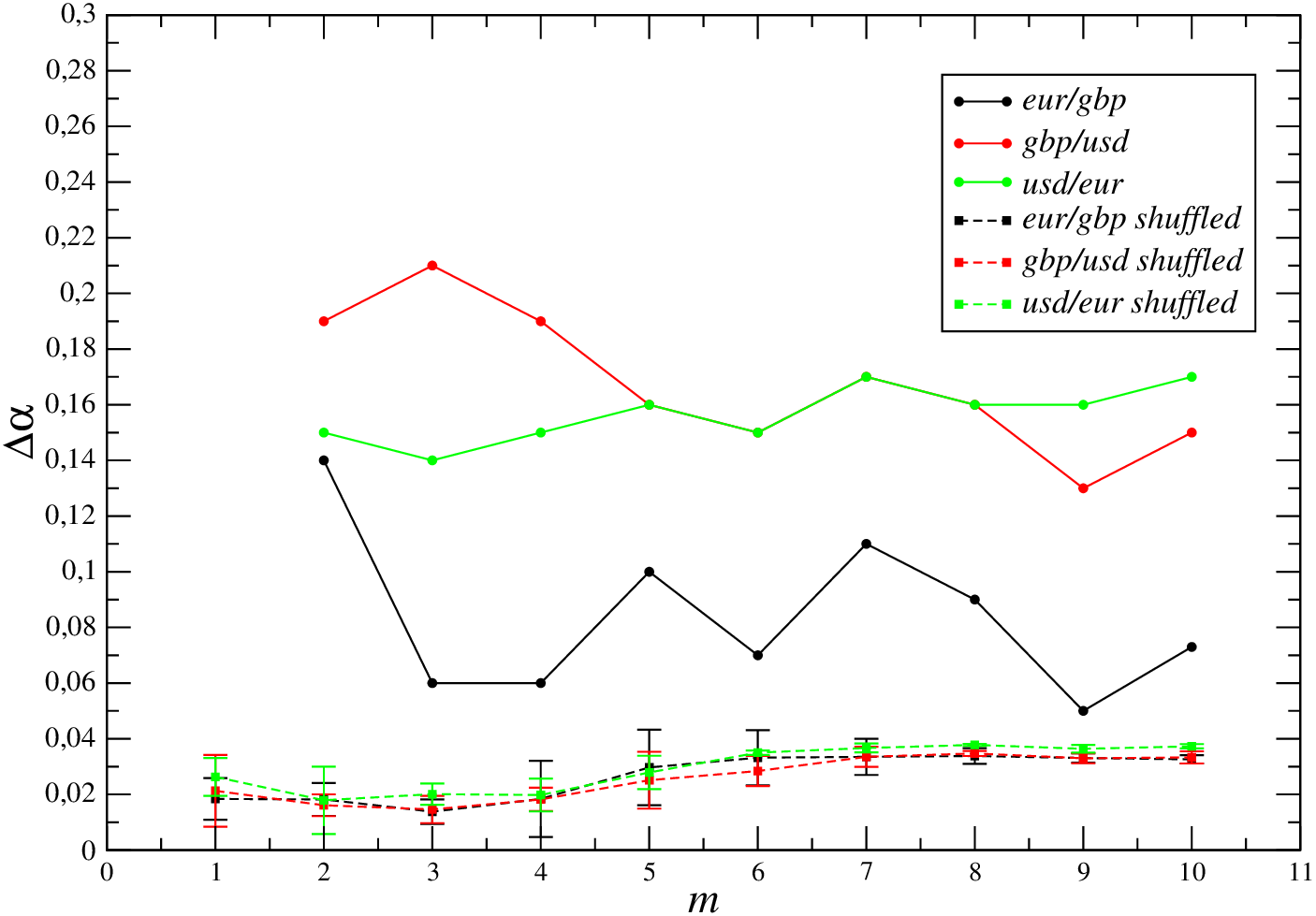}
\caption{Hurst exponent $H$ (left) and  width of the singularity spectrum $\Delta \alpha$ (right) as functions of the detrending polynomial order $m$ for the forex data. The solid and dotted lines refer to the original and the reshuffled data, respectively. Error bars indicate standard deviation calculated from 10 independent randomly shuffled time series.}
\label{figg7}
\end{figure}

Financial time series are an example of data generated by extremely complicated processes. The power law distribution of fluctuations of such signals and their non-trivial structure of correlations causes that analysis of such data requires often applying advanced methods of analysis. In our study, we focus on the data coming from the foreign exchange market. We consider one-minute logarithmic returns of the following three exchange rates: EUR/GBP, GBP/USD, USD/EUR, quoted over the period from January 2, 2004 to March 30, 2008 ~\cite{drozdz10}. It is known that temporal correlations of such type of signals can be quantified through the multifractal characteristics. Moreover, we study the exchange rates from the triangle GBP-EUR-USD, which possibly include triangular arbitrage opportunities. In our approach, we focus on the Hurst exponent and the widths of the singularity spectra. In Figure \ref{figg7} (left panel), we depict the $H$ as a function of the number of the detrending polynomial degree $m$. It is clear that, for the so small $m$, the time series reveal weak antipersitence. For $m>3$ the singularity specta shift to the left and, consequently, the value of the Hurst index decreases for all the considered currencies. The rate of decreasing $H(m)$ is similar for all signals, whereas the smallest $H$ is obtained for the EUR/GBP exchange rate: $H(10)=0.39$. In the same Figure, we also show the average outcomes for the reshuffled data. In this case, $H$ calculated for $m<3$ is approximately equal to $0.5$ indicating uncorrelated data. For larger values of $m$, the Hurst index for the reshuffled data decreases, similarly to $H$ for the original ones. However, the pace of this decrease is slower in the former case. This suggests that, for the analyzed data, the correlations affect effectiveness of the detrending procedure, especially in the case of large $m$. In Figure \ref{figg7} (right panel), we present the estimated $\Delta \alpha$ as a function of $m$ for the original as well as for the reshuffled data. As we expected, $f(\alpha)$ for the original data are much wider then for the randomly shuffled ones. Moreover, we can see that the character of the $\Delta \alpha$ function depends on the analyzed exchange rate. For the pair GBP/USD, the width of $f(\alpha)$ is a decreasing function of $m$, whereas for EUR/GBP and USD/EUR, $\Delta \alpha$ is rather stable. In the case of the reshuffled data, the considered characteristics is similar for all the time series. The shape of this function is also similar to the corresponding characteristics obtained for the FBM process with minimum at $m=3$ and weakly increasing $\Delta \alpha$ for larger values of $m$.

\section{Analysis of literary texts}

\begin{figure}
\begin{center}
\includegraphics[width=0.65\textwidth,height=0.5\textwidth]{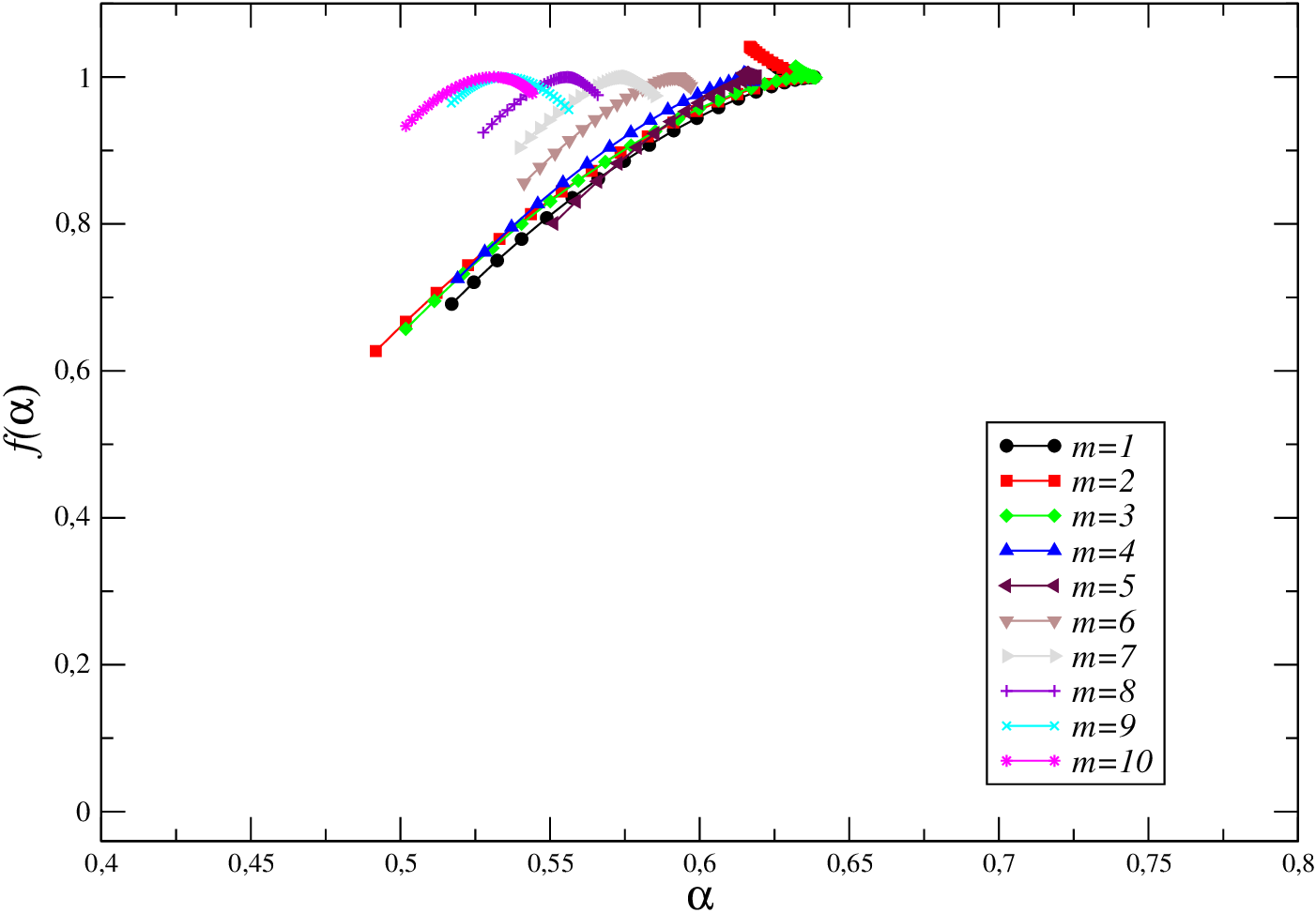}
\end{center}
\vspace{0.5cm}
\begin{center}
\includegraphics[width=0.65\textwidth,height=0.5\textwidth]{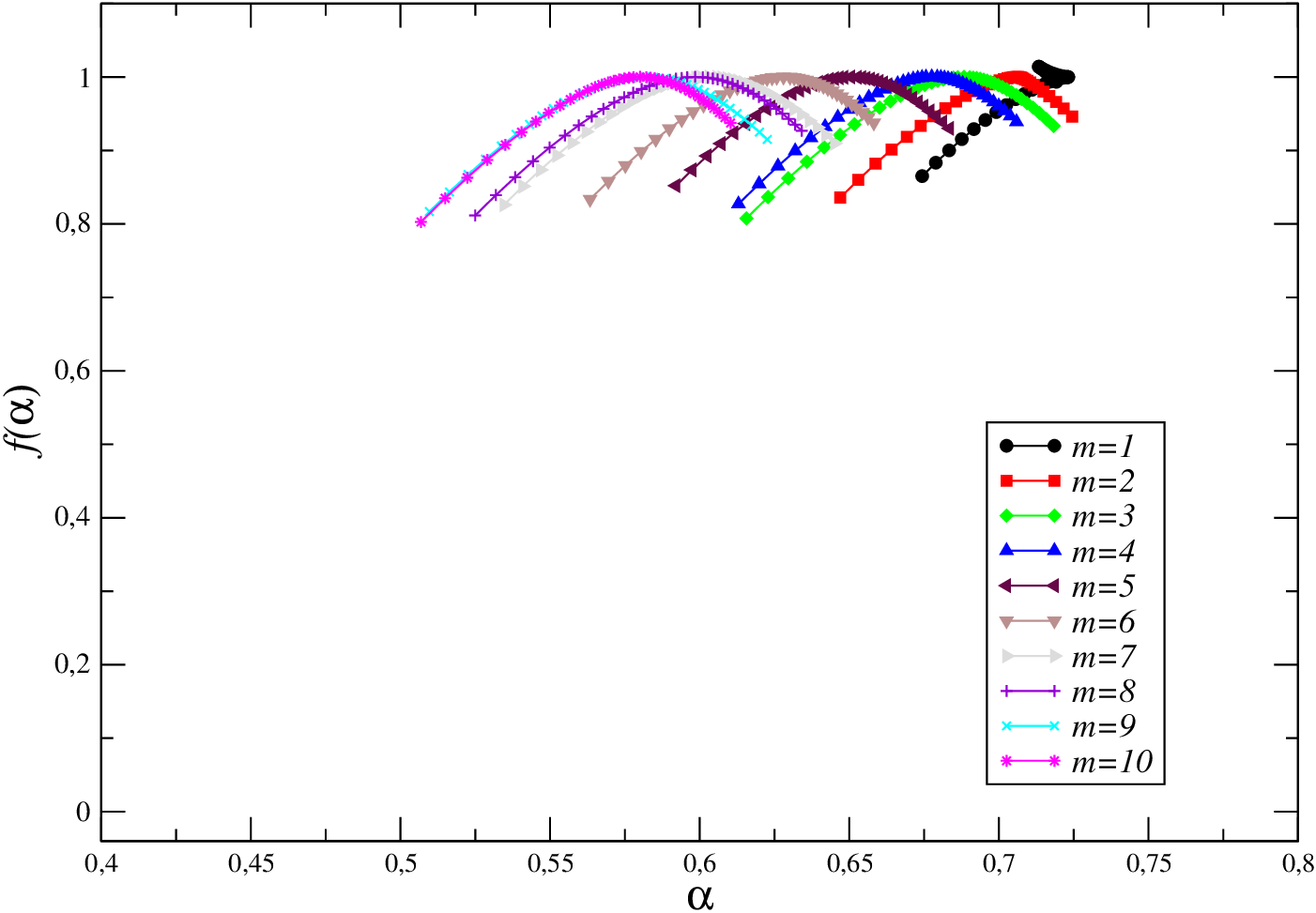}
\end{center}
\caption{Singularity spectra $f(\alpha)$ for the texts: "Alice's Adventures in Wonderland" (top) and "Moby Dick" (bottom). Different detrending polynomials in MFDFA are used.}
\label{figg8}

\end{figure}

In recent years, literary texts have attracted growing attention of scientists from such fields of science like, for
example, physics, mathematics, and computer science ~\cite{christiansen08,kwapien2012,ausloos12}. On the one hand, the
reason for this is an opportunity of an insight into principles of the human brain's information processing and into
structure and evolution of natural language. On the other hand, however, the ability to transform the written texts into
a series of numbers gives a chance of applying the advanced methods of time series analysis to explore the statistical
and dynamical properties of language. In our study, we consider two English literaly texts: "The Alice's Adventures in
Wonderland" by Lewis Carrol and "Moby Dick; or, the Whale" by Herman Melville. At first, we transform these texts into
numerical representation according to the following rules. For both books, the frequency of words occurrence is
estimated by dividing the number of appearances of each word by the total number of words (the text's length). Next,
each word is replaced with its rank which is attributed according to the ordered frequency table of the words in the
text. So, the most frequently appearing word receives the rank one, next in a row gets the rank two, and so on. In the
end, we obtained two time series which now can be analyzed by means of MFDFA. In Figure \ref{figg8}, we depict
$f(\alpha)$ for both considered texts, calculated for different orders of the detrending polynomial.  

\begin{figure}
\includegraphics[width=0.5\textwidth,height=0.4\textwidth]{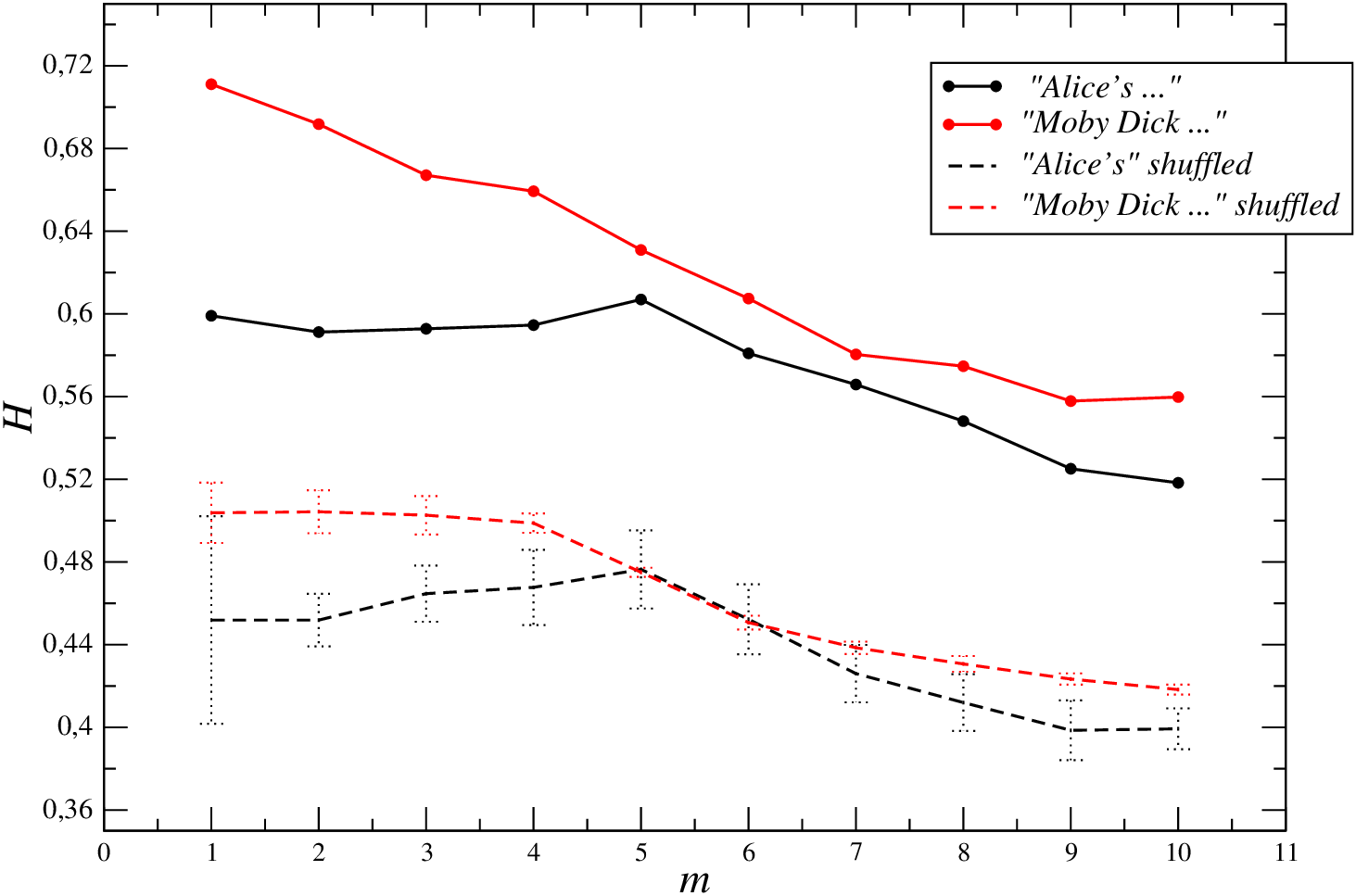}
\hspace{0.3cm}
\includegraphics[width=0.5\textwidth,height=0.4\textwidth]{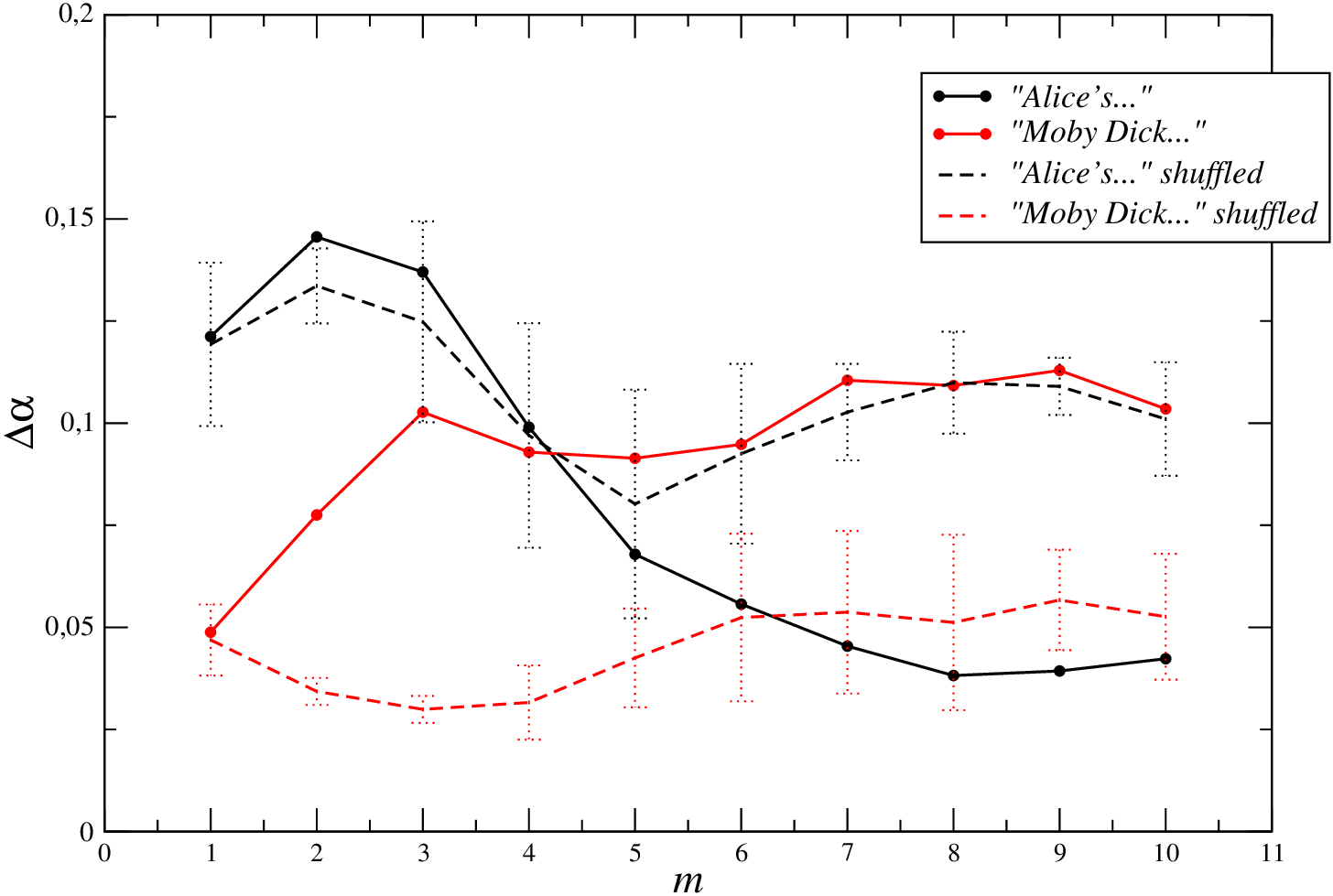}
\caption{The Hurst exponent $H$ (left) and the width of the singularity spectrum $\Delta \alpha$ (right) as functions of the detrending polynomial order $m$, calculated for "Alice's.. " and for "Moby Dick". Solid and dotted lines refer to the original and the reshuffled data, respectively.  Error bars indicate standard deviation calculated from 10 independent randomly shuffled time series.}
\label{figg9}
\end{figure}

The singularity spectra estimated for each of the books differ from each other. In the case of Carrol's book, 
the spectra calculated for small $m$'s are wide and asymmetric with evidently longer left wing. However, the larger is
the degree of MFDFA, the narrower and more symmetric is the singularity spectrum. Asymmetric multifractal spectra are
also obtained for small $m$'s in the case of $Moby Dick$, but in contrast to the previous text, the larger is the value
of $m$, the wider is the $f (\alpha)$ function. For both books, with the decrease of $m$, the maximum of the
multifractal spectra moves towards the smaller values of $\alpha$. This result is presented quantitatively in Figure
\ref{figg9}. The Hurst exponents estimated for small $m$'s indicate strong persistence for "Moby Dick" ($H(2)=0.69$) and
a bit weaker linear correlations for "Alice's..." ($H(2)=0.6$). Moreover, in the former case, $H(m)$ is a monotonically
decreasing function for all the considered values of $m$, while in the latter case, the Hurst index is approximately
constant for $m<6$ and decreases only for larger $m$. In the same Figure, we show the average Hurst exponents calculated
for ten independent reshufflings of the data. In this case, the form of the $H(m)$ function is similar to its
counterpart for the  Forex data. The only difference is that, for the text of "Alice's...", the Hurst exponent weakly
increases with $m$ in the range of $1<m<6$. In the right panel of Figure \ref{figg9}, the widths of the multifractal
specra are presented as a function of the parameter $m$. It is clearly visible that the estimated $\Delta \alpha$ are
small enough to allow us to consider the investigated signals monofractals. Interestingly, the behaviour of $\Delta
\alpha (m)$ is different for each time series. For "Alice's...", the width of $f(\alpha)$ decreases with $m$ and is even
smaller than its counterpart for the randomly shuffled data. However, $\Delta \alpha$ derived for the second text looks
completely different. The singularity spectrum's width strongly increases for small $m$ and weakly fluctuates around the
value of $0.1$ for $m>3$. The corresponding quantity for the shuffled data takes the value approximately equal to
$0.05$.

\section{Conclusions}

We investigated the effect of using different detrending polynomials on the
multifractal spectra obtained with the MFDFA technique. This issue is crucial
for correct characterization of dynamics of the considered signals. In the
case of real-word signals, true form of a trend is not known \textit{a
priori}. Approximating a trend by e.g. random polynomial carries a risk of
incorrect calculation of multifractal spectra. Therefore, so important is to
address the question of stability of the results for different variants of
MFDFA. We considered artificially generated fractals with {\it a priori} known
properties as well as some real-world signals with unknown fractal
characteristics. We showed that the mutlifractal spectrum is shifted towards
the antipersistent regime with increasing the detrending polynomial order. The
speed of this movement of $f(\alpha)$ depends on the correlations present in
the analyzed time series. We thus suspect that this kind of characteristics
could provide additional information about temporal and spatial dependencies
present in time
series and indicate true form of the trend present in data.

The detrending procedure affects also the richness of multifractality. In the
case of mathematical mono- and multifractals, the higher is the order of the
assumed polynomial trend, the stronger multifractality is uncovered. Even
though $\Delta \alpha$ of a bi-fractal is insensitive to the order of a
detrending polynomial, in the case of real signals, typically, one does not
have a unique model of the behaviour of the $\Delta \alpha(m)$ function. For
example, for "Alice's...", the estimated complexity (fractality) of the time
series decreases with the increase of the polynomial order, but for "Moby
Dick", $\Delta \alpha(m)$ is a non-decreasing function. A similar situation is
observed for the Forex data, where the effect of detrending on the richness of
multifractality is characteristic for individual exchange rates.

These outcomes show that one of the basic multifractal characteristics i.e.
width of the singularity spectrum also depends on assumed form of a trend. In
practice, within the MFDFA method, a polynomial of order two is the most
frequently used. However, as the results for binomial cascade suggest, in some
cases, a polynomial of higher order can give us more accurate fractal
characteristics. Moreover, different kinds of trend could describe different
parts of a signal. This situation appears for time series with variable
$\Delta \alpha(m)$, being the effect of more effective detrending of one type
of fluctuations. These findings suggest that MFDFA can give us much more
information about the correlation structure of data than does a multifractal
spectrum alone. Thus, research in this subject should be continued in future.

\end{document}